# Surface magnetic anisotropy at a compensated interface of ferromagnetic-antiferromagnetic bilayer

N. A. Usov
*Troitsk Institute for Innovation and Fusion Research, 142190, Troitsk, Moscow region, Russia*

Ching-Ray Chang and Zung-Hang Wei
*Department of Physics, National Taiwan University, Taipei, Taiwan, 10764, Republic of China*

The periodic deviations of ferromagnetic and antiferromagnetic spins from corresponding uniform configurations are shown to be energetically favorable close to a compensated interface of ferromagnetic – antiferromagnetic bilayer. The amplitude of the deviations decreases exponentially into AFM and FM volumes as a function of a coordinate perpendicular to the interface. The interaction energy is found to be proportional to a square of a scalar product of unit ferromagnetic and antiferromagnetic vectors.

The exchange coupling is observed in many magnetic multilayers where ferromagnetic (FM) and antiferromagnetic (AFM) thin films are in atomic contact.[1-3] This phenomenon is a consequence of a fundamental interaction between the AFM and FM spins at the interface. Recently, some important features of the AFM-FM exchange coupling have been successfully studied both experimentally and theoretically.[2-10] In particular, the numerical simulations of Koon[6] showed the possibility of so-called perpendicular coupling at a compensated AFM surface, which is expected to have no net interfacial moment. The existence of the perpendicular coupling at the compensated AFM-FM interface was confirmed also in numerical calculations of Schulthess and Butler[7,8], though they stated this type of coupling did not necessarily lead to the exchange bias. These theoretical predictions are in agreement with the experiments on FeF$_2$/Fe and MnF$_2$/Fe bilayers[11-13] with (110) and (101) compensated orientations at the interface. The perpendicular coupling is also confirmed for Fe$_3$O$_4$/CoO bilayers with compensated (001) CoO interface.[14-16] Nevertheless, the nature of the phenomenon is still not understood completely.

In this paper we prove that certain periodic modulation of the AFM and FM spin structure near a compensated surface of AFM is energetically favorable. The calculations are based on Heisenberg Hamiltonian for an AFM-FM bilayer. The amplitude of the modulation is shown to decrease exponentially into AFM and FM volumes as a function of a coordinate perpendicular to the interface. Therefore, from a macroscopic point of view the AFM-FM exchange interaction can be considered as a surface magnetic anisotropy. It is shown that the interaction energy is proportional to the square of a scalar product of unit FM and AFM vectors. As a result, the surface magnetic anisotropy is invariant with respect to arbitrary rotation of the AFM-FM spin system as a whole. The surface anisotropy constant, $K_s$, is determined as a function of Heisenberg Hamiltonian parameters. Due to condition $K_s > 0$, the corresponding effective interaction makes the FM spins rotate perpendicular to the direction of the AFM spins at the interface.

Consider a system of AFM spins, $q_{n,m,k}$, and FM spins, $s_{n,m,k}$, interacting on a simple cubic lattice ($n$, $m$, $k$). Let upper $N_f$ plates, $k = 1, 2, .. N_f$, contain FM spins, whereas lower $N_a$ plates, $k = -N_a+1, -N_a+2, .. 0$, contain AFM spins, respectively. Neglecting first magnetic anisotropy, Heisenberg Hamiltonian of this system is given by

$$H = H_f + H_a + H_{\text{int}}, \qquad (1)$$

where

$$H_f = -J_f \sum_{n,m} \sum_{k=1}^{N_f} s_{n,m,k} \left( s_{n+1,m,k} + s_{n,m+1,k} + s_{n,m,k+1} \right);$$

$$H_a = J_a \sum_{n,m} \sum_{k=0}^{-N_a+1} q_{n,m,k} \left( q_{n+1,m,k} + q_{n,m+1,k} + q_{n,m,k-1} \right);$$

and

$$H_{\text{int}} = J_{\text{int}} \sum_{n,m} s_{n,m,1} q_{n,m,0} .$$

Here $J_f > 0$ and $J_a > 0$ are the FM and AFM exchange integrals, respectively, $J_{\text{int}}$ being the exchange integral for the AFM-FM interaction. In the mean-field approximation[17] a naïve solution to the Hamiltonian (1) is given by

$$s_{n,m,k} = S\alpha ; \quad q_{n,m,k} = Q(-1)^{n+m+k} \beta, \qquad (2)$$

where $\alpha$ and $\beta$ are arbitrary unit vectors, $|\alpha| = |\beta| = 1$, $S$ and $Q$ are the lengths of FM and AFM spins, respectively. For this solution the total energy of the Hamiltonian (1) is easily calculated as

$$E_0 = -3J_f S^2 N(N_f - 1) - 3J_a Q^2 N(N_a - 1), \qquad (3)$$

where $N \gg 1$ is the number of the lattice sites ($n$, $m$) at the interface plane.

It is important to note, that the energy $E_0$ does not depend on the directions of vectors $\alpha$ and $\beta$, as well as on the interaction strength $J_{\text{int}}$. One can suspect therefore, that a solution with a lower energy can be found for the Hamiltonian (1). The latter has to take into account the interaction of FM and AFM spins at the interface. To construct this solution one can use a well-known fact[17] that Hamiltonian (1) is invariant with respect to arbitrary 3 dimensional (3D) rotations. Therefore, without loss of generality, one can assume that vector $\boldsymbol{\beta}$ is parallel to the $x$ axis, whereas vector $\boldsymbol{\alpha}$ makes an angle $\varphi$ with this axis, being parallel to the '$xy$' plane for simplicity. With this assumption in mind, let us consider the following trial solution

$$\mathbf{s}_{n,m,k} = S\{\cos(\varphi + (-1)^{n+m}u(k)), \sin(\varphi + (-1)^{n+m}u(k)), 0\} \quad (4a)$$

for $k = 1, 2, .., N_f$, and

$$\mathbf{q}_{n,m,k} = Q\{(-1)^{n+m+k}\sqrt{1-v^2(k)}; (-1)^k v(k); 0\} \quad (4b)$$

for $k = -N_a+1, -N_a+2, .., 0$, where the perturbations $u(k)$ and $v(k)$ are assumed to be small enough. The Hamiltonian (1) can be evaluated as a function of the perturbations up to the second order in these parameters. The result is given by $H = E_0 + H_1$, where

$$\frac{H_1}{N} = J_f S^2 \sum_{k=1}^{N_f}\left[4u^2(k) + \frac{1}{2}[u(k)-u(k+1)]^2\right] +$$

$$J_a Q^2 \sum_{k=0}^{-N_a+1}\left[4v^2(k) + \frac{1}{2}[v(k)-v(k-1)]^2\right] +$$

$$J_{int}\sin\varphi QS[v(0)-u(1)]. \quad (5)$$

We will prove below that the perturbations $u(k)$ and $v(k)$ decrease exponentially as functions of $|k|$. Therefore, in the first approximation one can take into account in Eq. (5) only perturbations $u(1)$ and $v(0)$. Then the first and second sums in Eq. (5) can be replaced by $4.5u^2(1)$ and $4.5v^2(0)$, respectively. In this approximation the minimum of the $H_1$ corresponds to the value

$$H_1 = -N\frac{J_f S^2 + J_a Q^2}{18 J_f J_a} J_{int}^2 \sin^2\varphi, \quad (6a)$$

the equilibrium perturbations being

$$u_0(1) = \frac{QJ_{int}}{9SJ_f}\sin\varphi; \quad v_0(0) = -\frac{SJ_{int}}{9QJ_a}\sin\varphi. \quad (6b)$$

Consider now the minimum of the first sum in Eq. (5) at a given value of the perturbation $u(1)$. Making a variation of this sum with respect to arbitrary deviations $u(k) \to u(k) + \delta u(k)$, one can obtain a recurrence relation for the equilibrium perturbations

$$u(k-1) - 10u(k) + u(k+1) = 0; \quad k = 2, 3, ..\ .$$

The solution of this relation is given by $u(k) = u(1)t^{k-1}$, where $t = 5 - 24^{1/2} \approx 0.1$. This means that the perturbations $u(k)$ decrease exponentially as function of $k$. Therefore, in the limit $N_f \gg 1$ the equilibrium value of the first sum in Eq. (5) is given by

$$\sum_{k=1}^{N_f}\left[4u^2(k) + \frac{1}{2}[u(k)-u(k+1)]^2\right] =$$

$$u^2(1)\frac{4+0.5(1-t)^2}{1-t^2} \approx 4.4u^2(1).$$

Similar calculation shows that the second sum in Eq. (5) is approximately given by $4.4v^2(0)$. Both values are very close to those ones used above. With this correction in mind, one can obtain from Eq. (6a) that the interaction energy per unit square of the AFM-FM interface is given by

$$E(\varphi) = -K_s \sin^2\varphi, \quad (7a)$$

where

$$K_s = \frac{J_f S^2 + J_a Q^2}{17.6 J_f J_a a^2} J_{int}^2. \quad (7b)$$

Here $a$ is the lattice period and $K_s$ has the meaning of a surface anisotropy constant.

To illustrate the nature of the solution (4), let us consider a simple case when $S = Q = 1$, $J_f = J_a = 100k_B$, where $k_B$ is the Boltzmann's constant, and $a = 4\cdot10^{-8}$ cm. Then numerically, the surface anisotropy constant $K_s \approx (J_{int}/J_f)^2$ erg/cm$^2$. If $J_{int} < J_f$, say $J_{int} = 30k_B$, one has $K_s = 0.1$ erg/cm$^2$. For this case the amplitude of the deviations (6b) is small, so that the solution (4) is very close to the naïve solution (2), as Fig.1 demonstrates for the particular angle $\varphi = \pi/2$. In Fig. 1, as well as in other figures only FM plate $k = 1$ and AFM plate $k = 0$ are shown, because the amplitude of the periodic deviations of the unit FM and AFM vectors turns out to be negligibly small for other plates.

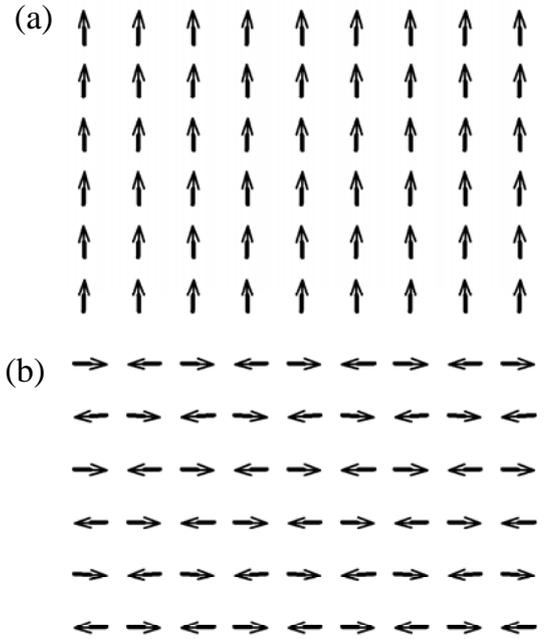

FIG. 1. Lowest energy state in the case of small anisotropy constant, $K_s = 0.1$ erg/cm$^2$: a) FM plate, $k = 1$; b) AFM plate, $k = 0$.

On the other hand, if $J_{int} \geq J_f$, for example $J_{int} = 200 k_B$, one has $K_s = 4$ erg/cm$^2$. For this case the deviations (6b) are appreciable. They are maximal when the average direction of the unit FM vector is perpendicular to that of the unit AFM vector, $\varphi = \pi/2$. This situation is shown in Fig. 2. Note, at this angle the interaction energy (7) has the lowest value, $E(\pi/2) = -K_s$. However, the amplitudes $u_0(1)$ and $v_0(0)$ diminish if angle $\varphi$ decreases, so that they become zero at $\varphi = 0$. To demonstrate the angle dependence of the spin structure (4) the spin arrangements in the FM and AFM layers closest to the interface are shown in Fig. 3 for $\varphi = \pi/4$.

It follows from Eq. (5) that the orientation deviations of spins in both the AFM and FM films near the AFM-FM interface are energetically favorable. Actually, the energy gain (6a) decreases if we assume, for example, the FM film to be strictly uniform, $u(k) = 0$. It is clear that dipolar and anisotropy energy contributions will decrease the amplitude of the variations (6b). But these contributions are small with respect to the exchange interaction energy (1) and can be considered as perturbations. Recently[18], the numerical simulation based on self-consistent mean-field approximation for Heisenberg Hamiltonian[19] has been carried out for AFM-FM bilayer with a compensated interface, the dipolar and anisotropy energy contributions being taken into account.

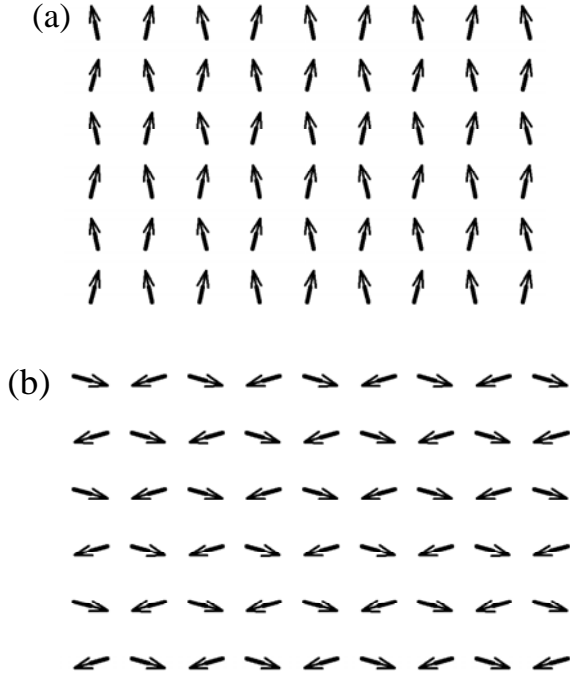

FIG. 2. FM and AFM spin structures for the case of high anisotropy constant, $K_s = 4$ erg/cm$^2$, $\varphi = \pi/2$: a) FM plate, $k = 1$; b) AFM plate, $k = 0$.

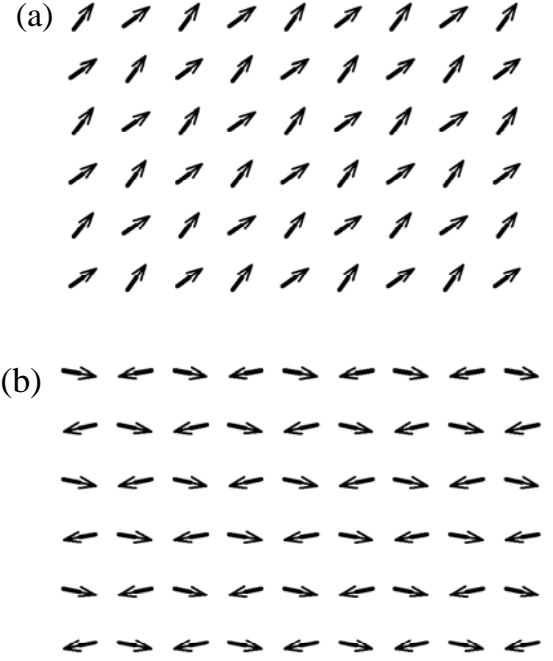

FIG. 3. The same as in Fig 2, but for $\varphi = \pi/4$.

This calculation confirms the existence of the periodic modulation of the spin structure near the compensated interface described by Eq. (4).

It is instructive to rewrite AFM-FM interaction energy density (7) as $E(\alpha,\beta) = K_s(\alpha\beta)^2$, where unimportant constant is omitted. This expression shows explicitly that the interaction energy is invariant with respect to an arbitrary 3D rotation of the AFM-FM system as a whole. Actually, one can easily check that the following generalization of the trial solution (4) for arbitrary directions of $\alpha$ and $\beta$

$$s_{n,m,k} = S\left\{\alpha\sqrt{1-u^2(k)} + (-1)^{n+m} u(k) \frac{[\alpha \times [\alpha \times \beta]]}{|[\alpha \times \beta]|}\right\}$$

for $k = 1, 2, .. N_f$, and

$$q_{n,m,k} = Q\left\{(-1)^{n+m+k}\beta\sqrt{1-v^2(k)} + (-1)^k v(k) \frac{[\beta \times [\alpha \times \beta]]}{|[\alpha \times \beta]|}\right\}$$

for $k = -N_a+1, -N_a+2, .. 0$, leads to the same expression (5) provided that $\varphi$ is the angle between the vectors $\alpha$ and $\beta$.

Note that the spin arrangement (4) corresponds to a compensated AFM-FM interface. Qualitatively, it can describe exchange interaction at the (110) and (101) compensated planes of FeF$_2$/Fe and MnF$_2$/Fe bilayers. It follows from Eq. (7) that for a compensated AFM-FM interface the lowest energy state corresponds to the case when FM spins turn out to be perpendicular to the AFM ones. In other words, perpendicular coupling is preferable in this case, in agreement with the experimental findings.[11-13]

Another type of spin arrangement at the AFM surface corresponds to that of (001) plane of single crystals of CoO or NiO.[14-16] In this case the AFM spins are parallel to each other within the (111) planes, but have opposite directions for adjacent (111) planes. Therefore, at the (001) CoO plane the spin structure looks like an array of alternating lines of spins.[14] Qualitatively, one can model this situation assuming the following expression for the Hamiltonian $H_a$ in Eq. (1)

$$H_a = -J_{ax} \sum_{n,m} \sum_{k=0}^{-N_a+1} \mathbf{q}_{n,m,k} \left( \mathbf{q}_{n+1,m,k} + \mathbf{q}_{n,m,k-1} \right) + J_{ay} \sum_{n,m} \sum_{k=0}^{-N_a+1} \mathbf{q}_{n,m,k} \mathbf{q}_{n,m+1,k} . \quad (8)$$

For this AFM Hamiltonian equilibrium spin structure corresponds to a sequence of ferromagnetic planes (010) with opposite directions of spins for adjacent planes. Therefore, the naïve solution for total Hamiltonian is now

$$\mathbf{s}_{n,m,k} = S\boldsymbol{\alpha}; \qquad \mathbf{q}_{n,m,k} = Q(-1)^m \boldsymbol{\beta} .$$

Here $\boldsymbol{\alpha}$ and $\boldsymbol{\beta}$ are arbitrary unit vectors. The total energy for this solution is given by

$$E_0 = -3J_f S^2 N(N_f - 1) - (2J_{ax} + J_{ay}) Q^2 N(N_a - 1) .$$

Using the fact that the total Hamiltonian is rotationally invariant, without loss of generality one can consider a trial solution of the type

$$\mathbf{s}_{n,m,k} = S\left\{ \cos(\varphi + (-1)^m u(k)); \sin(\varphi + (-1)^m u(k)); 0 \right\} \quad (9a)$$

for $k = 1, 2, .., N_f$, and

$$\mathbf{q}_{n,m,k} = Q\left\{ (-1)^m \sqrt{1 - v^2(k)}; v(k); 0 \right\} \quad (9b)$$

for $k = -N_a+1, -N_a+2, .., 0$. The total energy corresponding to the trial solution (9) is given by $H = E_0 + H_1$, where

$$\frac{H_1}{N} = J_f S^2 \sum_{k=1}^{N_f} \left[ 2u^2(k) + \frac{1}{2}[u(k) - u(k+1)]^2 \right] +$$

$$\frac{1}{2} J_{ax} Q^2 \sum_{k=0}^{-N_a+1} [v(k) - v(k-1)]^2 + 2J_{ay} Q^2 \sum_{k=0}^{-N_a+1} v^2(k) +$$

$$J_{int} \sin\varphi \, QS[v(0) - u(1)]. \quad (10)$$

Minimizing the first sum in Eq. (10) at a given value of the perturbation $u(1)$ one obtains a recurrence relation

$$u(k-1) - 6u(k) + u(k+1) = 0$$

for equilibrium perturbations. This leads to the conclusion that $u(k) = u(1) \, t^{k-1}$, where $t = 3 - 8^{1/2} \approx 1/6$. Therefore, perturbations $u(k)$ decrease exponentially into FM volume and the first term in Eq. (10) is estimated to be $2.33 J_f S^2 u^2(1)$. One also obtains similar estimation, $2.33 J_a Q^2 v^2(0)$, for the sum of the second and third terms in Eq. (10) if one assumes for simplicity that $J_{ax} = J_{ay} = J_a$. As a result, we arrive to the same Eq. (7) for the interaction energy with the difference that the numerical coefficient 17.6 in the expression for $K_s$ is corrected for 9.3.

Using the values $J_f = 16 \cdot \text{meV}$, $J_a = 1.2 \cdot \text{meV}$ for the exchange integrals[20,21], $S = 1$, $Q = 2$ for the spin lengths[22] and $a \approx 4 \cdot 10^{-8}$ cm for the lattice period in FeF$_2$ [12] one can estimate the surface anisotropy constant of bilayer FeF$_2$/Fe as $K_s \approx 0.1 (J_{int}/J_a)^2$ erg/cm$^2$. To get the experimentally determined value[2] of the bilayer FeF$_2$/Fe interface energy, $\Delta E = 0.5 - 1.3$ erg/cm$^2$, it is sufficient to assume that unknown value of the interfacial exchange integral $J_{int}$ is somewhere in between the $J_a$ and $J_f$ values. Similarly, assuming all exchange integrals for bilayer Fe$_3$O$_4$/CoO to be of the same order of magnitude, $J_{ex} \approx 2.5 \cdot 10^{-15}$ erg[14], and using $a = 4.2 \cdot 10^{-8}$ cm for CoO[16], one can estimate the surface anisotropy constant for this bilayer as $K_s \approx 1$ erg/cm$^2$, in reasonable agreement with the experimental value[2], $\Delta E = 1.43$ erg/cm$^2$.

On the other hand, in most of the cases studied[2,3,11-16] the experimental situation is more complicated because the AFM surface consists of twin domains with different directions of easy anisotropy axes. The concept of surface magnetic anisotropy can be applied to this case too, the results of the calculations will be presented elsewhere.[18]